\documentclass[a4paper,11pt]{article}
\usepackage{arxiv}
\usepackage{amsmath,amsfonts,amssymb}
\usepackage{graphicx}
\usepackage{setspace}
\usepackage{tocloft}
\usepackage{lineno}
\usepackage{natbib}

\cftpagenumbersoff{figure}
\cftpagenumbersoff{table} 

\usepackage[colorlinks=true, allcolors=blue]{hyperref}
\usepackage{threeparttable} 
\usepackage{macro_ref}

\title{High-angular resolution and high-contrast VLTI observations from Y to L band with the Asgard instrumental suite} 
\renewcommand{\shorttitle}{\textit{High-angular resolution and high-contrast observations with the Asgard instrumental suite}}

\date{}

\begin{document} 
\maketitle
\vspace{-6em}
\begin{center}
Marc-Antoine Martinod$^{a}$\footnote{marc-antoine.martinod@kuleuven.be},
Denis Defrère$^{a}$,
Michael Ireland$^{b}$,
Stefan Kraus$^{c}$,
Frantz Martinache$^{d}$,
\newline
Peter Tuthill$^{e}$,
Azzurra Bigioli$^{a}$,
Julia Bryant$^{e}$,
Sorabh Chhabra$^{c}$,
Benjamin Courtney-Barrer$^{b}$,
\newline
Fred Crous$^{e}$,
Nick Cvetojevic$^{d}$,
Colin Dandumont$^{f}$,
Germain Garreau$^{a}$,
Tiphaine Lagadec$^{b}$,
\newline
Romain Laugier$^{a}$,
Daniel Mortimer$^{c}$,
Barnaby Norris$^{e}$,
Gordon Robertson$^{e}$,
Adam Taras$^{e}$
~\newline

$^a$Institute of Astronomy, KU Leuven, Celestijnenlaan 200D, 3001 Leuven, Belgium\\
$^b$The Australian National University, Canberra, Australia\\
$^c$University of Exeter, School of Physics and Astronomy, Stocker Road, Exeter, United Kingdom\\
$^d$Laboratoire Lagrange, Observatoire de la Côte d'Azur, Nice, France\\
$^e$Sydney Institute for Astronomy, School of Physics,  University of Sydney, NSW 2006, Australia\\
$^f$Space sciences, Technologies \& Astrophysics Research (STAR) Institute, University of Li\`ege, Li\`ege, Belgium
\end{center}
~\vspace{-0.5em}

\begin{abstract}
The Very Large Telescope Interferometer is one of the most proficient observatories in the world for high angular resolution.
Since its first observations, it has hosted several interferometric instruments operating in various bandwidths in the infrared. As a result, the VLTI has yielded countless discoveries and technological breakthroughs.
Here, we introduce a new concept for the VLTI, Asgard: an instrumental suite comprised of four natively collaborating instruments: BIFROST, a combiner whose main science case is studying the formation processes and properties of stellar and planetary systems; NOTT, a nulling interferometer dedicated to imaging young nearby planetary systems in the L band; HEIMDALLR, an all-in-one instrument performing both fringe tracking and stellar interferometry with the same optics; Baldr, a Strehl optimiser. 
These instruments share common goals and technologies.
The goals are diverse astrophysical cases such as the study of the formation and evolution processes of binary systems, exoplanetary systems and protoplanetary disks, the characterization of orbital parameters and spin-orbit alignment of multiple systems, the characterization of the exoplanets, and the study of exozodiacal disks.
Thus, the idea of this suite is to make the instruments interoperable and complementary to deliver unprecedented sensitivity and accuracy from the J to M bands to meet these goals. 
The interoperability of the Asgard instruments and their integration in the VLTI are major challenges for this project.

\end{abstract}

\keywords{integrated-optics, wavefront control, infrared, high contrast imaging, high angular resolution, optical fibers, long baseline interferometry, exoplanets, AGN, protoplanetary disk, spectroscopy}



\section{Introduction}
\label{sec:intro}  
The emphatic triumph of the Very Large Telescope Interferometer (VLTI) and its second-generation instruments in delivering unique science has set European astronomy apart, motivating the ongoing facility upgrade within the Gravity+ framework  \citep{Gravityplus2019} promising still further ground-breaking scientific discoveries. 
In parallel, major technology and scientific milestones have been achieved on other interferometric facilities, such as the Center for High Angular Resolution Astronomy Array (CHARA) and the Large Binocular Telescope Interferometer (LBTI). 
New ideas and laboratory demonstrations have also emerged in recent years opening and enabling the path to novel scientific capabilities for optical interferometry. 
Leveraging these recent developments, the Asgard instrument suite \citep{asgard_martinod} will extend the scientific capabilities of the VLTI with a set of four instrument modules: BIFROST \citep{bifrost_kraus, bifrost_chhabra, bifrost_mortimer} (Beam-combination Instrument for studying the Formation and fundamental paRameters of Stars and planeTary systems) which is a Y/J/H-band combiner optimized for high spectral resolution, HEIMDALLR \citep{ireland2018} (High-Efficiency Multiaxial Do-it ALL Recombiner) which is a high-sensitivity K-band fringe tracker, NOTT \citep{hi5_defrere, hi5_dandumontA, hi5_dandumontB, hi5_garreau, hi5_laugier, Laugier2022, hi5_sanny} (Nulling Observations of dusT and planeTs), formerly known as Hi-5 \citep{Defrere2018}, which is an L-band nuller optimized for high-contrast observations, and Baldr which is an H-band injection optimizer for BIFROST and NOTT.

This paper presents the instruments, capabilities and key scientific objectives of Asgard. 
The first step in bringing the Asgard suite to the VLTI is to propose it as a visitor instrument, serving as a critical new platform for test and launch of future innovations to deliver benefit to a broad community for years to come. 

\section{Science cases for Asgard}

Asgard science cases mainly revolve around stellar physics, multiple systems and exoplanetary science \citep{bifrost_kraus, hi5_defrere}.

\paragraph{Formation process of binary systems:}
BIFROST and HEIMDALLR will aim to spatially resolve and completely characterize up to 6000 binaries among the 28 million non-single stars in the GAIA DR3 catalog \citep{bifrost_kraus}.
Both instruments will measure visibilities and closure phases.
They will derive different information about multiple systems from these observables.
They will find the statistics of orbital parameters by solving the degeneracy of the GAIA measurement of the photocenter. 
Combined with statistics coming from radial velocity observations and direct imaging, this data will help to constrain the formation process of binary systems.
These instruments will also allow the derivation of the dynamical masses, providing an input for the refinement of evolutionary models that are the cornerstone of modern astrophysics, as these models currently face uncertainties, for instance  the treatment of mass-loss in massive stars \citep{Feiden2012, Mann2015}. 
The uncertainties are particularly notable in the pre-main-sequence phase, where predicted and measured masses currently differ by $\geq 10\%$ \citep{Stassun2014}.
HEIMDALLR and BIFROST will measure the binary flux ratio in the K-band and Y/J-band respectively, where BIFROST's operating wavelengths are close to the GAIA band (0.4-0.9 $\mu$m). 
This will allow deriving dynamical masses with a precision of $\sim3\%$.
The mass constraints will also give access to the ages of the stars, which is important for Galactic Archaeology studies.
Finally, BIFROST will give information about the processes involved in the formation and dynamical evolution of binaries and planetary systems by measuring the spin-orbit alignment of the binaries and planet-host stars. 
This is enabled through BIFROST’s very high-spectral resolution (VHR) mode with R=25,000 ($\Delta v \sim 12$~km/s), which allows measuring the stellar spin orientation in moderate/slow rotating stars.
The statistics obtained from these observations will enable tests of theories concerning the origin of this obliquity, such as multiple modes of formation, the Kozai-Lidov mechanism or the flyby of neighboring stars.

\paragraph{Mass accretion and ejection:}
BIFROST will also have a particular interest in studying the process of mass accretion and ejection from Young Stellar Objects (YSO) and Active Galactic Nuclei (AGN).
To do so, the instrument will perform differential visibility measurements in the Paschen lines and the  He I 1.083~$\mu$m line; these are stronger markers than Br$\gamma$ lines in the K band \citep{Nicholls2017}.
Thus, it will gather critical information on accretion disks, wind-wind collisions, disk winds and relativistic jets from micro-quasars.
In addition, by spatially and spectrally resolving multiple transitions of hydrogen recombination lines, BIFROST can constrain the physical conditions (gas density, temperature, excitation) and the velocity field at different regions in the circumstellar environment. 
This will enable new approaches for reconstruction of the 3-dimensional velocity field in the inner regions of protoplanetary disks, where accretion occurs and winds are launched from the star, the disk, or the interaction region between the stellar and disk magnetic fields.

\paragraph{Formation and evolution of exoplanetary systems and exoplanet atmospheres:}
Although the GAIA mission provides useful information about the masses of giant gas exoplanets and the ages of the systems, it only brings a partial knowledge about the formation and evolution of exoplanetary systems.
NOTT will fill the gap by measuring the size and the effective temperatures of the exoplanets in order to better constrain their planet-formation models \citep{wallace_likelihood_2019}.

NOTT's spectroscopic capability will enable us to trace the thermal emission from these exoplanets, opacities of absorbers (molecular bands, dust, clouds) and BIFROST will be able to probe the the exoplanet atmospheres using its off-axis mode, in particular those of Jupiter-like exoplanets.
They will give access to critical information about the non-equilibrium chemistry of their atmosphere, their vertical temperature profiles, atmospheric dynamics, rotation rates and formation processes.

\paragraph{Protoplanetary and circumplanetary disks:}
Protoplanetary disks are structured with the presence of rings, gaps, spirals and large cavities that provide clues about the formation process.
The observed cavities are often large ($>20$~AU) and are observed in 10\% of the disks  \citep{vanderMarel2018} and many of them show asymmetric features like spirals or vortices \citep{vanderMarel2021} that are linked to the presence of companions carving the cavities in the disk \citep{Norfolk2021}.
NOTT observations will provide unprecendented inputs for hydrodynamical models to understand planet-disk interaction and constrain planet formation and evolution in the very early stages.
The high-contrast nulling mode of NOTT combined with the long VLTI baselines will provide the dynamic range and angular resolution required to observe the cavities of transition disks imaged at shorter wavelengths by current direct imaging instruments such as SPHERE, and currently inaccessible to other L-band (single-dish) exoplanet imagers.

BIFROST's high-resolution spectroscopic capability will probe the circumplanetary disks where exoplanets are formed.
The instrument will investigate the Pa$\gamma$ and Pa$\beta$ lines in the circumplanetary disk, thereby revealing the kinematics in the circumplanetary disk and the accretion onto the forming planets.
BIFROST offers up to 6 times higher spectral resolution than GRAVITY and opens a spectral window of particular interest for line studies.

\paragraph{Exozodiacal dust}
Exozodiacal dust is a warm and hot dust in the inner regions of main-sequence planetary systems, including the Habitable Zone (HZ).
It presents both a scientific interest and an obstacle for imaging a future exo-Earth. 
Indeed, the light from HZ dust adds photon noise and causes confusion in the discrimination of disk structure   \citep{defrere_direct_2012}.
Exozodiacal dust has so far only been observed in thermal emission due to the lower dust-to-star contrast compared to scattered light, but the dust’s excess emission is still too faint to detect photometrically.
Thus, it needs to be spatially resolved from the star, and the small angular scales and high contrasts involved (1 AU at 10 parsec corresponds to 0.1”) require the use of precision interferometry.
Particularly, nulling interferometry on the VLTI with NOTT will provide tenfold better sensitivity to extended circumstellar emission than the stellar optical long-baseline interferometers like PIONIER \citep{lebouquin2011} and FLUOR \citep{Coude2003, Scott2013}.
NOTT will enable image reconstruction and thus detailed study of the origin of the dust through its distribution, in addition to the previously mentioned structures showing planet-disk interaction.

\section{Instrument Capabilities}
Asgard will extend the scientific capabilities of the VLTI in many ways: shorter wavelengths combined with high-spectral resolution, high-sensitivity and high-performance fringe tracking, and high-contrast direct imaging with the first VLTI nulling instrument.
To do so, Asgard relies on the  following four instruments:
\begin{itemize}
    \item BIFROST will offer dramatic benefits compared to the 2nd-generation VLTI instruments, namely a better angular resolution (due to shorter wavelengths), a higher continuum flux for blue sources (as we observe closer to the peak wavelength of the photosphere), around 6 times higher spectral resolution (R=25,000), and access to new line tracers, including Pa$\beta$ 1.282 $\mu$m, Pa$\gamma$ 1.094 $\mu$m, the He~I~1.083 $\mu$m accretion-tracing line \citep{Fischer2008}, and forbidden lines (e.g. [Fe II] 1.257 $\mu$m). To do so, BIFROST will have two arms: one with a low resolution (LR) spectrograph and the other one with a high resolution (HR) spectrograph  \citep{bifrost_kraus}.
    
    \item HEIMDALLR is a dual-band and multiaxial beam combiner, where beams are placed in a two-dimensional array in a non-redundant layout, then combined in the manner analogous to  an aperture-masking interferometer. This innovation delivers: low order wavefront control by the retention of the full two-dimensional information contained in the interferogram, high sensitivity by the use of bulk optics, and accurate closure-phase using the proven concept of aperture masking interferometry.

    \item Baldr uses the same detector as HEIMDALLR, and provides a Zernike Wavefront Sensor or a photonic lantern to augment NAOMI (in number of modes and speed) or GPAO (Gravity+ AO, in speed for low-order modes). This system determines the practical magnitude limit for BIFROST as a secondary adaptive optics system is needed to increase short-wavelength Strehl. It can operate in either J or H bands, leaving the rest of Y,J,H for BIFROST science.

    \item NOTT will be the first nulling instrument at the VLTI, the first long-baseline nuller in the Southern hemisphere, and the first operating at L band, where young giant planets are the most luminous. Leveraging the long-baseline of the VLTI and the ongoing Gravity+ facility upgrade, it will be able to directly image the snow line where most giant exoplanets are located \citep{Fulton2021}. It relies on integrated optics for beam combination in order to provide a compact and stable design \citep{hi5_sanny}. In addition, NOTT will be able to compensate the fringe drift due to water vapor and CO$_2$ dispersion.
\end{itemize}

The main instrumental features of the VLTI Asgard science modules are given in Table~\ref{tab:features} and the preliminary performances are given in Table~\ref{tab:performances}.
\begin{table}[h]
\centering
\renewcommand{\arraystretch}{1.25} 
\begin{threeparttable}[b]
    \centering
    \caption{Main instrumental features of the VLTI Asgard modules}
    \begin{tabular}{|c|c|c|c|c|}
    \hline
        \textbf{Feature} & \textbf{BIFROST} & \textbf{HEIMDALLR} & \textbf{Baldr} & \textbf{NOTT}\\
    \hline
        Photometric band & Y, J, H & K & H & L\\
    \hline
    Central wavelength ($\mu$m) & 1.35  & 2.18 & 1.6 & 3.75 \\
    \hline
        Bandwidth ($\mu$m) & 0.6 & 0.45 & 0.3 & 0.5 \\
    \hline
        Spectral resolutions& \begin{tabular}{@{}c@{}}
                                \textbf{LR arm}: R=50 \\ 
                                \textbf{HR arm} \\
                                MR: R=1000 \\ (1.05-1.65 $\mu$m) \\
                                HR: R=5000 \\ (1.05-1.65 $\mu$m)\\
                                VHR: R=25,000 \\
                                (around He~I+Pa$\beta$ \\ and Pa$\gamma$)\end{tabular} & R=5 (2 channels) & None &
                                \begin{tabular}{@{}c@{}}
                                R = 20\\
                                R = 400\\
                                R = 2000\tnote{$^\ast$}\\
                                    \end{tabular}\\
    \hline
        Polarization split & YES (optional) & NO & - & YES\tnote{$^\ast$}\\
    \hline
        Off-axis mode & YES\tnote{$^\ast$} & - & - & - \\
    \hline
        \begin{tabular}{@{}c@{}}
        Inner Working Angle \\
        @3.75 $\mu$m
        \end{tabular} & - & - & - & \begin{tabular}{@{}c@{}} 2 mas (AT),\\ 
                                                        3 mas (UT)\end{tabular}\\
    \hline
        \begin{tabular}{@{}c@{}}
        Diameter of the \\
        field of view\end{tabular} & 
        \begin{tabular}{@{}c@{}}
        155 mas (AT),\\ 34 mas (UT) \\ @1.35 $\mu$m \end{tabular}  &
        \begin{tabular}{@{}c@{}}
        222 mas (AT),\\ 49 mas (UT) \\@ 1.94 $\mu$m \end{tabular} &
        \begin{tabular}{@{}c@{}}
        183 mas (AT),\\ 41 mas (UT) \\@ 1.6 $\mu$m \end{tabular} &
        \begin{tabular}{@{}c@{}}
        430 mas (AT),\\ 94 mas (UT) \\ @3.75 $\mu$m \end{tabular} \\
    \hline
       \begin{tabular}{@{}c@{}}
        Interferometric \\
        observable\end{tabular} &
       \begin{tabular}{@{}c@{}}
        V$^2$,
        closure phase, \\
        differential phases\end{tabular} &
        V$^2$,
        closure phase &
        - &
       \begin{tabular}{@{}c@{}}
        Null depth, \\ differential \\ null depth \end{tabular}\\        
    \hline
        Magnitude limit &
        Limited by Baldr &
        \begin{tabular}{@{}c@{}}
        K=11.5 (ATs) or \\
        13.5 (UTs)\end{tabular} &
        \begin{tabular}{@{}c@{}}
        H=9.6 (ATs) or \\ 12.9 (UTs) \end{tabular} &
        \begin{tabular}{@{}c@{}}
        L’=11 (ATs) or \\
        13 (UTs) \end{tabular}\\
    \hline
    \end{tabular}
    \label{tab:features}

    \begin{tablenotes}
       \item[$^\ast$]Contingent on additional funding.
     \end{tablenotes}
\end{threeparttable}
\end{table}

\begin{table}[h]
\centering
\renewcommand{\arraystretch}{1.25} 
\begin{threeparttable}[b]
    \centering
    \caption{Preliminary performance of the VLTI Asgard modules}
    \begin{tabular}{|c|c|c|c|c|}
    \hline
        \textbf{Feature} & \textbf{BIFROST} & \textbf{HEIMDALLR} & \textbf{Baldr} & \textbf{NOTT}\\
    \hline
    Precision on $V^2$ & 3\% (LR arm) & 5\% & - & -\\
    \hline
    Precision on closure phase & $0.5^\circ$ (LR arm) & $1^\circ$ & - & -\\
    \hline
    Precision on differential phase & $0.1^\circ$ (HR arm) & - & - & -\\
    \hline
    \begin{tabular}{@{}c@{}}
        Residual piston RMS (nm)
        \end{tabular} & - & \begin{tabular}{@{}c@{}} $< 100$ (UT)\\ $< 50$ (AT) \end{tabular} & - & \begin{tabular}{@{}c@{}} $< 100$ (UT)\\ $< 50$ (AT) \end{tabular}\\
    \hline
        Strehl (in J band) & - & - & \begin{tabular}{@{}c@{}} $> 50$\% (AT) \\ $> 40$\% (UT)\end{tabular} & -\\
    \hline
    \begin{tabular}{@{}c@{}}
        RMS intensity fluctuations\\ (in L band)
        \end{tabular} & - & - & $< 2$\% & -\\
    \hline
    \begin{tabular}{@{}c@{}}
        Precision on null depth \\
        (L=4 \& R=18)\end{tabular} & - & - & - & \begin{tabular}{@{}c@{}} $5 \times 10^{-5}$ (UT)\\ $3 \times 10^{-4}$ (AT) \end{tabular}\\
    \hline
    \end{tabular}
    \label{tab:performances}
\end{threeparttable}
\end{table}

\section{Instrumental overview}
\subsection{Instrumental setup and components}
\begin{figure}[h]
    \centering
    \includegraphics[width=0.8\textwidth]{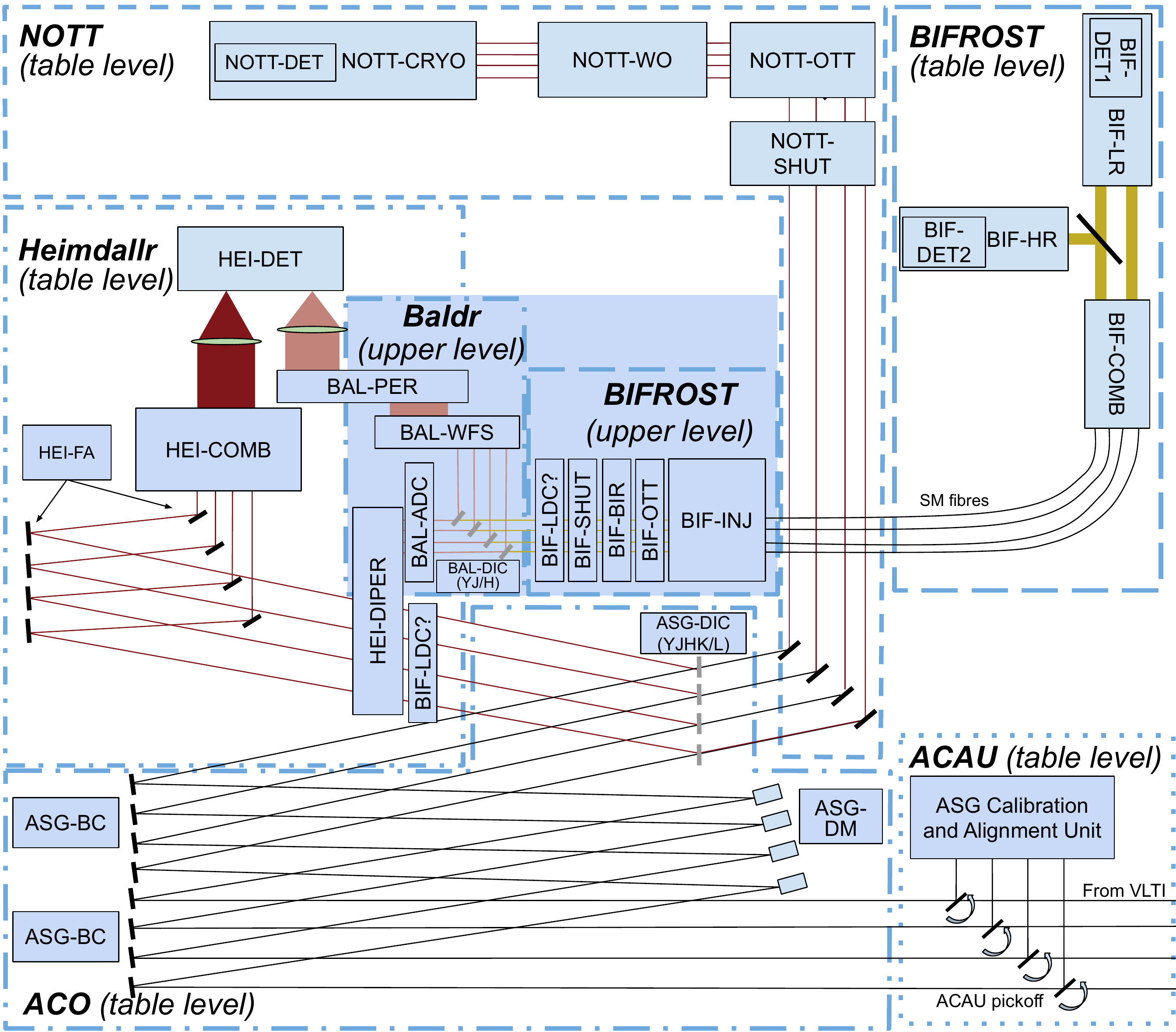}
    \caption{Conceptual layout of Asgard. This diagram is not to scale and the positions of the modules are subject to change but the order of the optical components should remain.}
    \label{fig:asgard_diagram}
\end{figure}

Asgard is a suite of four instruments that will be located on the Visitor 2 table (formerly the AMBER table). 
It will operate from Y to L bands (1 to 4 microns). The conceptual layout of Asgard is given in Figure~\ref{fig:asgard_diagram}.
The four 18-mm beams from the VLTI (bottom right) are first reflected on the beam compressors, which reduce the beam size to 12 mm using off-axis paraboloids.
The deformable mirrors are located in the pupil plane between the two sets of mirrors of the beam compressors.
The L band light is then sent to NOTT with a dichroic while the Y-K band light is sent to the other modules. 
More information on the sub-systems of each module is given below. 

Asgard has two common modules for all the instruments. The first is the Asgard Calibration and Alignment Unit (ACAU), which contains a light source for spectral calibration and a 4-beam coherent light source for internal alignment and cophasing. 
It also provides pickoffs that are flippable mirrors to select between light from ACAU and the VLTI.

The second common module is the Asgard Common Optics (ACO) which includes the following subsystems:
\begin{itemize}
    \item Beam compressors (ASG-BC): consist of 4 pairs of optics to reduce the beams’ size from 18 mm to 12 mm;
    \item Deformable mirrors (ASG-DM): comprise 4 deformable mirrors (DM) actuated by MEMS technology. They perform fringe tracking and injection control as well as the correction of higher aberration modes to maximise injection into the optical fibers of BIFROST and NOTT;
    \item Dichroic (ASG-DIC): transmitting the L band to NOTT and reflecting the shorter wavelengths to the other modules.
\end{itemize}

HEIMDALLR has the following sub-systems:
\begin{itemize}
    \item Dichroic and periscope (HEI-DIPER): the dichroic separates the Y-J-H and K bands. The periscope sends the Y-J-H band to Baldr/BIF-INJ at the upper level. The K band is transmitted toward HEIMDALLR’s optics;
    \item Focus and alignment mirrors (HEI-FA): comprise 4 pairs of mirrors. They align and cophase HEIMDALLR. The mirrors are moved by stepper motor actuators;
    \item Pupil reconfiguration (HEI-COMB): comprising knife-edge mirrors to reconfigure the beams into a two-dimensional pattern. The K band-beams are reconfigured according to a 2-dimensional pattern providing 6 non-redundant baselines. The fringe pattern is used to perform wavefront sensing and science;
    \item Camera (HEI-DET): comprises the cooling system, its electronics and the detector. It collects and records the fringe pattern of HEIMDALLR and the signal from Baldr.
\end{itemize}

Baldr’s sub-systems are:
\begin{itemize}
    \item Atmospheric Dispersion Compensation (BAL-ADC): comprises 4 pairs of prisms. These devices will compensate for the chromatic dispersion effect between the Y and H bands, which degrades the injection efficiency and induces differential phases. They are actuated by stepper-motorised rotary stages;
    \item Beam splitters (BAL-DIC): reflecting the H-band to Baldr and transmitting the J-Y band to BIFROST. An alternative beam splitter can instead be translated into place that reflects the J band to Baldr and transmits the H band to BIFROST;
    \item Zernike optics (BAL-WFS): is the phase sensor. It will either be a wavefront sensor and comprise phase masks and shutters to decompose the wavefront on a Zernike aberration base, or photonic lanterns \citep{LeonSaval2013, Norris2020, Norris2022};
    \item Periscope (BAL-PER): sends the signals from BAL-ZER to the HEIMDALLR’s camera.
\end{itemize}

BIFROST’s sub-systems are:
\begin{itemize}
    \item Longitudinal dispersion compensator (BIF-LDC): each beam contains an LDC mounted on a piezo-electric linear stage to compensate for atmospheric longitudinal dispersion. Its location is yet to be determined: either in the common path with HEIMDALLR or in the BIFROST path only. The choice will depend on manufacturing quality (it needs an anti-reflection coating surface quality at better than 1\%), and the available space, as we have to be able to remove the subsystem entirely when NOTT and HEIMDALLR are used.
    \item Shutters (BIF-SHUT): comprises four shutters. They can individually close each beam or all of them to perform the calibration tasks or protect the detectors from bright light while BIFROST is not operating;
    \item Birefringence correction (BIF-BIR): comprises LiNbO3 plates that compensate for the birefringence of the beam trains and the guided optics. The LiNbO3 plates are set on stepper-motorised rotary stages and are adjusted to maximise contrast on BIFROST;
    \item OPD \& TT alignment mirrors (BIF-OTT): to cophase the input beams between BIFROST and HEIMDALLR and to optimise the light injection into BIF-INJ. A Differential Delay Line mirror mounted on a linear stage is used to cophase BIFROST with Asgard. A Tip-Tilt fast steering mirror is used to actively optimise the coupling into the optical fibers;
    \item Injection module (BIF-INJ): comprises off-axis paraboloids (OAPs) and the fiber tips. The fiber tips receive the light to transport it to the combiner of BIFROST through single-mode fibers. The fiber tips are fixed at the focus of the OAPs;
    \item Combiner (BIF-COMB): comprises the combiner and a 90/10 beam splitter that can be moved in to record data simultaneously with BIFROST’s LR arm and HR arm. The combiner interferes the four beams of starlight;
    \item LR arm (BIF-LR): comprises imaging optics, a prism for low-spectral dispersion  at a fixed resolving power (R=50 to 80, precise value TBD) and a camera. Also, it contains a Wollaston prism on a translation stage that can be moved in to measure separate polarisation states to improve the visibility calibration. The LR arm measures OPD drift, dispersion and fringe jumps to feed a feedback loop on the LDC and the BIFROST DLLs. The camera is a SAPHIRA-based APD detector system;
    \item LR arm detector (BIF-DET1): comprises the detector, its electronics and cooling system. It collects the low-dispersion fringe pattern and the dispersed photometric channels;
    \item HR arm (BIF-HR): comprises imaging optics, a filter wheel with grisms for high spectral dispersion of the fringe pattern and a camera. It is used for measuring wavelength-differential quantities and closure phases;
    \item HR arm detector (BIF-DET2): comprises the detector, its electronics and cooling system. It collects the high-dispersion fringe pattern. The camera is an APD detector system.
\end{itemize}

NOTT’s sub-systems are:
\begin{itemize}
    \item Beam conditioner (NOTT-SHUT): comprises shutters and diaphragms. It shuts the beams individually for calibration purposes and equalises the beam intensities to optimise the null depth;
    \item OPD \& TT alignment mirrors (NOTT-OTT): to cophase the input beams between NOTT and Asgard and to optimise the light injection into the photonic chip. They are actuated by stepper motors;
    \item Warm optics (NOTT-WO): comprises LDCs, polarisation control optics, the slicer and alignment camera. The LDC compensates for the chromatic phase across the L band and the water vapour effects via a feedback loop. The polarisation control compensates for polarisation effects which reduce the instrumental null depth. The slicer overlaps the four telescope pupils on the cold stop at the entrance of the cryostat. The alignment camera is an Infratec ImageIR 5300;
    \item Cryostat (NOTT-CRYO): comprises a window, a cold stop, the photonic chip \citep{hi5_sanny} which combines the beams and creates photometric outputs, a wheel and imaging optics, the camera, the cryocooler and the vacuum system. The wheel contains grisms for different spectral resolutions. The chip is mounted on a 3-axis mount for alignment and focusing with NOTT-WO;
    \item Camera (NOTT-DET): comprises the detector and its electronics. It collects the spectrally-dispersed signal delivered by the photonic chip. It consists of Teledyne’s 5-micron Hawaii-2RG with SIDECAR cold electronics and Astroblank MACIE warm electronics;
\end{itemize}

\subsection{Opto-mechanical design}
\begin{figure}[htbp]
    \centering
    \includegraphics[width=1.4\textwidth, angle=-90]{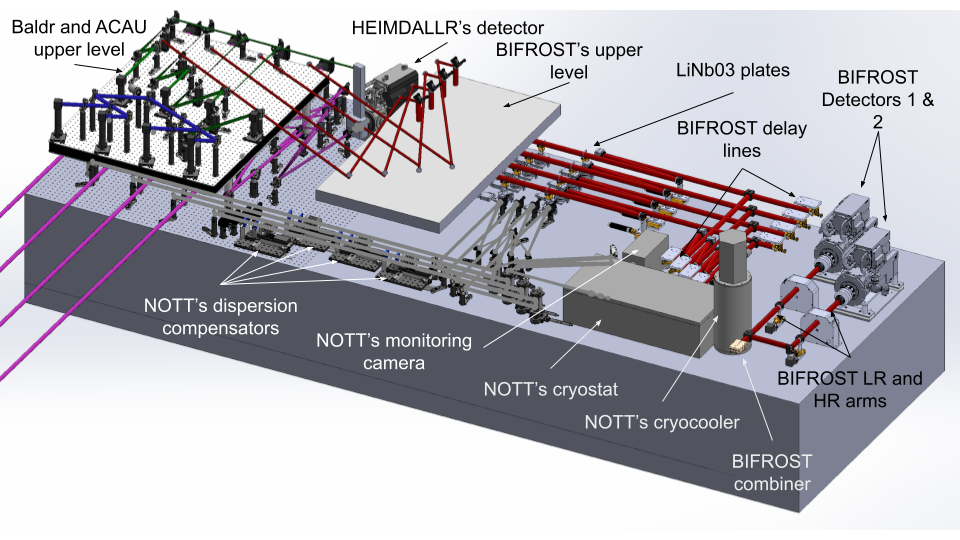}
    \caption{Preliminary version of the full opto-mechanical design of Asgard. The optical paths are colored for each module: HEIMDALLR (Purple), BIFROST (red), NOTT (gray), Baldr (green) and ACAU (Blue).}
    \label{fig:optomech}
\end{figure}

\begin{figure}[htbp]
    \centering
    \includegraphics[width=0.85\textwidth]{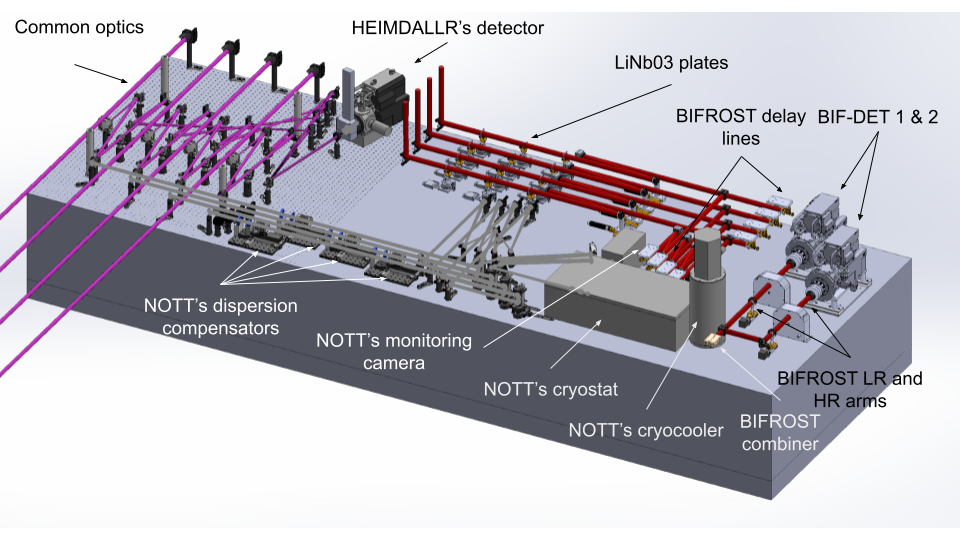} 
    \caption{Preliminary version of the opto-mechanical design of Asgard (optical table only). The optical paths are colored for each module: HEIMDALLR (Purple), BIFROST (red), NOTT (gray), Baldr (green) and ACAU (Blue).}
    \label{fig:optomech2}
\end{figure}

The conceptual layout (Fig.~\ref{fig:asgard_diagram}) has been developed into a preliminary opto-mechanical design to verify that Asgard can fit on the VLTI optical table (Fig.~\ref{fig:optomech} and \ref{fig:optomech2}).
The beams of the VLTI enter on the left hand side of the Visitor 2 table and come across ACO and HEIMDALLR's paths on the lower level.
A dichroic splits the light to send its L-band component to NOTT (gray beams) where they encounter dispersion correctors and the injection systems that are part of NOTT's warm optics and are then sent to the cryostat.
The remaining light is split again to send the K-band to HEIMDALLR's combiner and HEIMDALLR's detector while a periscope brings the YJH-band components to the upper level to be dispatched between Baldr (green beam) and BIFROST (red beams).
The red beams go along the visitor table, parallel to NOTT's beams.
They meet the delay lines before being injected into optical fibers (not shown) that carry them to the integrated-optical chip where they are coherently combined.
The beams of the coherent calibration source ACAU (blue beams) are created on the upper level and sent to the entrance of the table via periscopes.
This preliminary design demonstrates the possibility to fit 4 full scale instruments into a single optical table.

    

\subsection{Control architecture}
\begin{figure}[h]
    \centering
    \includegraphics[width=\textwidth]{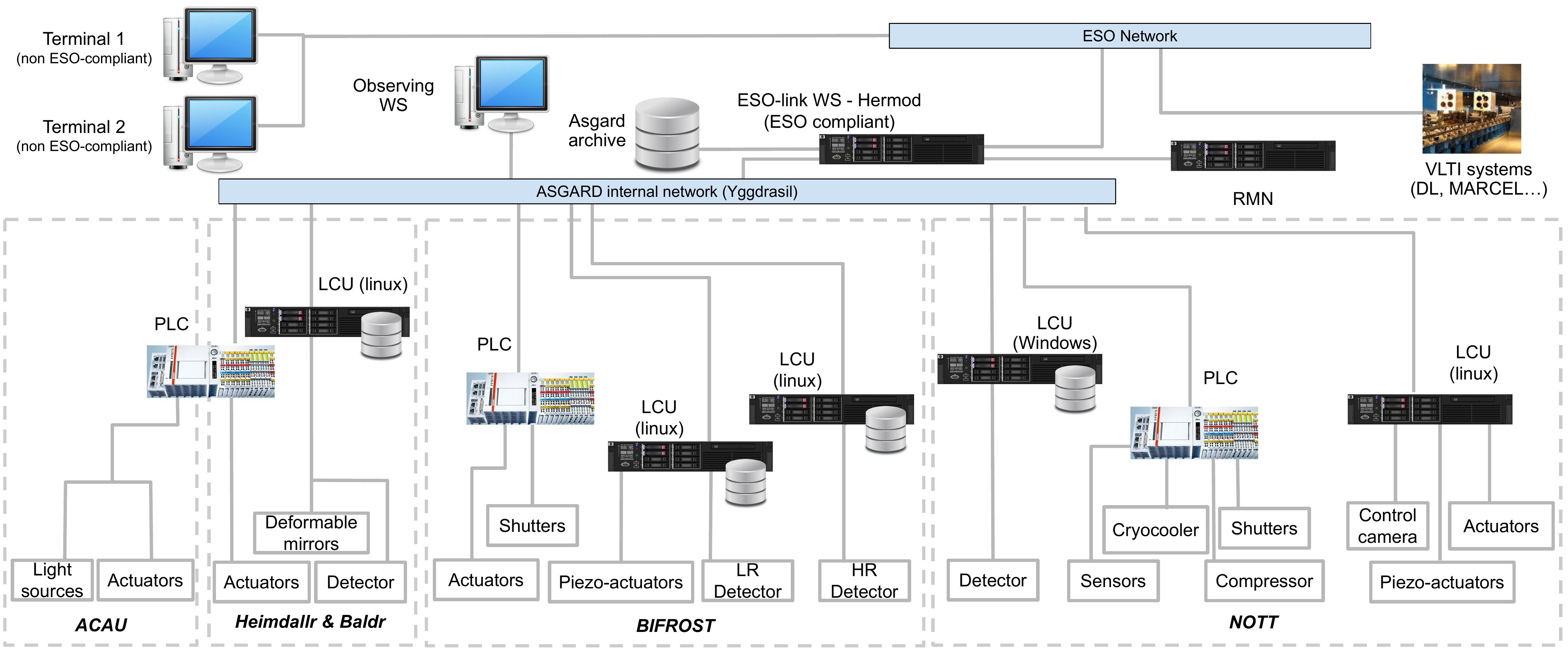}
    \caption{Preliminary Control architecture of Asgard.}
    \label{fig:asgard_architecture}
\end{figure}

The overall Asgard control architecture is given in Figure~\ref{fig:asgard_architecture}.
The hardware architecture is designed so that two terminals can operate all the modules.
This centralisation allows the observer to run all the instruments simultaneously and to communicate with the VLTI systems. 
These terminals can connect to the Observing Workstation (Observing WS) on Asgard through the ESO network while not necessarily being ESO compliant.

Asgard IT hardware and software are non-ESO compliant, hence must be kept separate.
However, Asgard modules need to interact with ESO systems such as the Reflective Memory Network (RMN), the delay lines and the calibration sources.
The suggested solution is to create a local network on which the IT components of each module are connected: the \emph{Asgard Internal Network} called \emph{Yggdrasil}\footnote{Sacred tree which supports all the worlds in the Nordic mythology}.
Asgard systems are run by the users through the Observing WS that is accessed either directly (during in-laboratory operations) or via the terminals (during the observations).
An \emph{ESO-link WS}, called \emph{Hermod}\footnote{Messenger of the gods in the Nordic mythology}, connected to \emph{Yggdrasil}, will interact with ESO systems, the RMN and transmit data to and from Asgard modules as required.
It is also connected to the \emph{Asgard Archive} to store Asgard data acquired over several weeks.
In a future extension, the Asgard archive could also serve as the hub for feeding Asgard data into the ESO archive.
\emph{Hermod} will respect ESO’s standards for hardware and software.
From the ESO network point of view, Asgard is a unique computer represented by \emph{Hermod}.

Each module has one or two Local Control Units (LCU) or Programmable Logic Controllers (PLC) to drive Asgard’s sub-systems and components including monitor cameras, science detectors and the piezo-actuators).
The LCUs are connected to the Asgard Internal Network and are driven from the terminals.
The PLCs directly control the opto-mechanical components and are controlled by the terminals similarly to the LCUs.
Each module (except ACAU) has its own data storage embedded in its respective LCU.

The LCUs managing the detectors have direct access to the archive to write the frames and the user can access the storage via the \emph{Observing WS}.
The connections between the WS and LCUs are made with Ethernet cables and the TTL protocol.\\

The Instrument Software Package will be subdivided into the following software modules:
\begin{itemize}
    \item The ICS (Instrument Control Software): to control all devices connected to the LCU on which it is installed, except the detector. There will be one ICS per module.
    \item The DCS (Detector Control Software): to control the detector. There will be one DCS per LCU connected to a detector.
    \item The OS (Observing Software): to coordinate the acquisitions of the instrument, interfaces with the \emph{\emph{ESO-link WS}}, with the local storage diskspace of the modules, and the Asgard archive.
    \item The LS (Link Software): exchanges information and instructions between ESO and Asgard networks.
\end{itemize}

\subsection{Data flow}
\begin{figure}[h]
    \centering
    \includegraphics[width=0.75\textwidth]{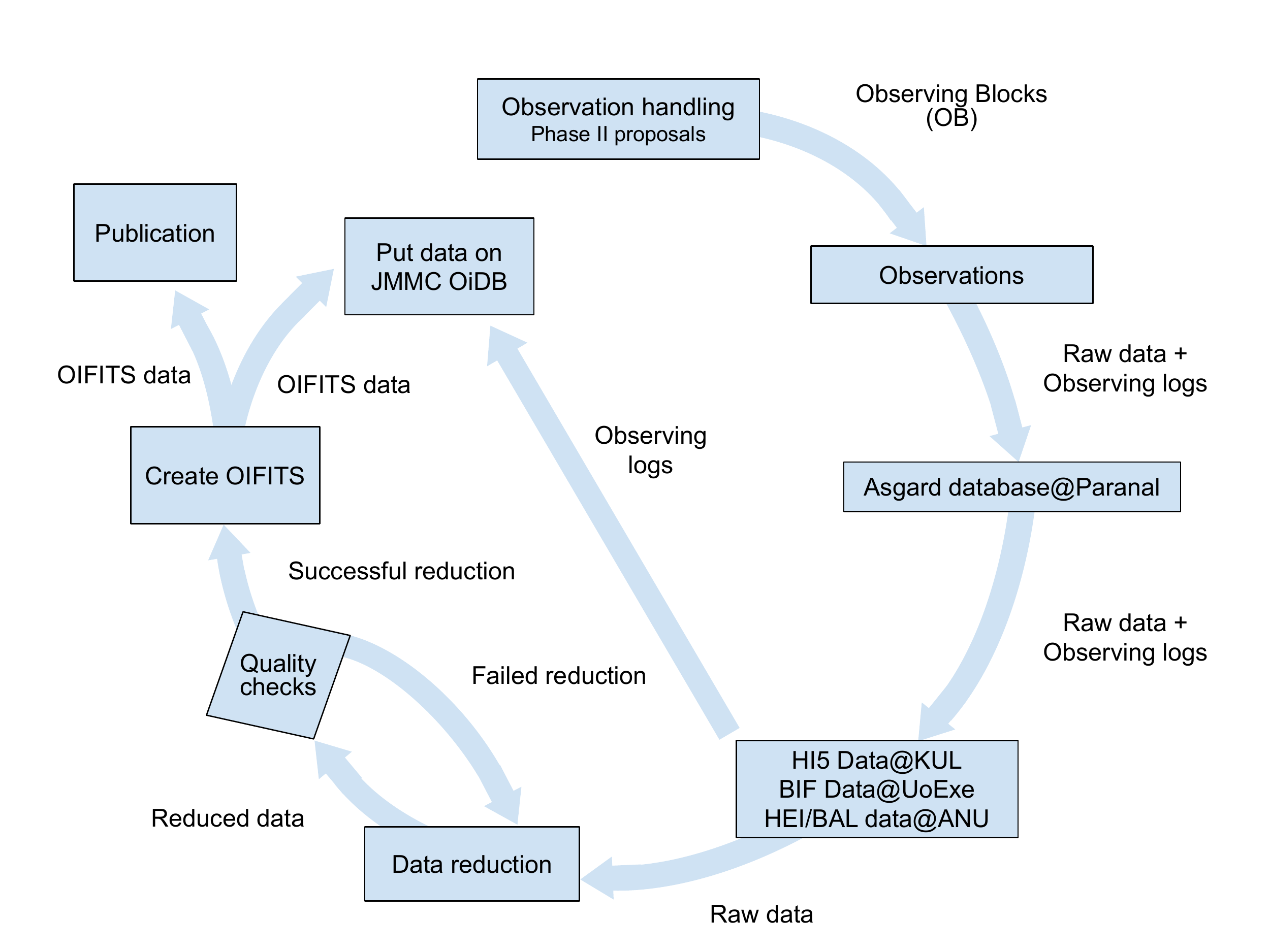}
    \caption{Preliminary data flow plan of Asgard.}
    \label{fig:dataflow}
\end{figure}

The data flow is shown in Figure~\ref{fig:dataflow}.
Asgard will use its own observing blocks.
After each observing night, the data (in FITS format) and the observing logs are copied to the local Asgard database. 
The data and the logs are copied to different places depending on the instrument (BIFROST data sent to the University of Exeter, HEIMDALLR/Baldr sent to ANU and NOTT data sent to KUL).
The transfer would be done by physically moving the storage drives. 
The data are reduced and quality controls check the quality of the final data. 
Reduced data are formatted in OIFITS files in order to be published and put on the JMMC OiDB with their observing logs.

\section{Summary}
Asgard is a new instrument suite proposed to ESO (P110, March 2022) to open new unique scientific capabilities at the VLTI (YJH high-spectral resolution and L-band high-contrast nulling interferometry) and to enable exoplanet imaging at milli-arcsecond angular separation.
It consists of 4 different modules: BIFROST (YJH high-spectral resolution (R=50, 1000, 5000, 25,000)), Baldr (Zernike Wavefront Sensor in H band),  HEIMDALLR (high-sensitivity fringe tracker (dual K band)), and NOTT (L-band (3.5-4 microns) nulling interferometer based on a photonic chip and self-calibration data reduction techniques).
The suite is currently in the design and preparation phases.
The design aims at making the instruments interoperable and making its data integrated to the existing data flows.

\section*{acknowledgments}
M-A.M. has received funding from the European Union’s Horizon 2020 research and innovation programme under grant agreement No.\ 101004719.\\
S.K. and S.C. acknowledge support from an ERC Consolidator Grant (``GAIA-BIFROST'', grant agreement No.\ 101003096).\\
S.K. and D.J.M. acknowledge support from STFC Consolidated Grant (ST/V000721/1).\\
SCIFY (A.B., C.D., D.D., G.G., and R.L.) has received funding from the European Research Council (ERC) under the European Union's Horizon 2020 research and innovation program (grant agreement CoG - 866070).
We are grateful for the kind support and constructive interactions with colleagues at ESO, in particular Frédéric Gonte, Xavier Haubois, Antoine Mérand, Nicolas Schuhler, and Julien Woillez.\\
This research was partially funded by the Australian Government through the Australian Research Council. (LE220100126).

\bibliography{report} 
\bibliographystyle{plainnat}

\end{document}